\documentclass[
    ,final            
  ]
  {aipproc}

\layoutstyle{6x9}

\begin{document}

\title{Measurement of reference cross sections \\ in pp and Pb--Pb collisions at the LHC \\ in van der Meer scans with the ALICE detector}



\author{M. Gagliardi, for the ALICE Collaboration}{
  address={INFN Sezione di Torino, V. Giuria 1, 10125 Torino, Italy - Martino.Gagliardi@cern.ch}
}




\begin{abstract}
Reference cross sections have been measured with the ALICE detector in both pp and Pb--Pb collisions, in dedicated luminosity calibration experiments (van der Meer scans) at the LHC. The results and their uncertainties are discussed, together with a few selected applications.
\end{abstract}

\maketitle


\section{Introduction}

The ALICE \cite{alice_JINST} experiment at the LHC \cite{lhc_JINST} aims at studying the behaviour of nuclear matter at high energy densities and the transition to the Quark Gluon Plasma (QGP), expected to occur in ultra-relativistic heavy ion collisions. 
Cross section measurements in pp collisions are essential for the ALICE physics program: particle production in heavy-ion (A--A) collisions is often compared with the extrapolation from elementary pp collisions via binary scaling (nuclear modification factor, $R_{\rm{AA}}$). 
The precision on $R_{\rm{AA}}$ needed  to quantify the importance of nuclear effects is typically $\simeq$10\%. Thus, a precision on the order of 5\% or better on the pp cross section (including luminosity normalisation) is desired in order not to be dominant in the overall uncertainty. 
Although it is not involved in $R_{\rm{AA}}$\footnote{The number of times a given process occurs per A--A collision at a given centrality is used instead.}, luminosity normalisation in Pb--Pb collisions is crucial for the measurement of other physics observables, e.g. in ultra-peripheral interactions. 


\section{vdM scan technique}

Measurement of the cross section $\sigma_{\rm{R}}$ for a chosen reference process is a prerequisite for luminosity normalisation: for a given data sample, the corresponding integrated luminosity can be obtained from the number of reference process events $N_{\rm{R}}$ in the sample as \hbox{$L_{\rm{int}}= N_{\rm{R}} / \sigma_{\rm{R}}$}.
Reference cross sections can be measured in  van der Meer (vdM) scans \cite{vdM}, where the two beams are moved one across the other in the transverse direction. Measurement of the rate $R$ of a given process as a function of the beam separation $\Delta X$, $\Delta Y$ allows one to determine the head-on luminosity $L$ as:
\begin{displaymath}
L = n N_1 N_2 f_{\rm{rev}} Q_{\rm{X}} Q_{\rm{Y}} ; \hspace{7 mm} Q_{\rm{X,Y}} = \frac{R(0,0)}{S_{\rm{X,Y}}}
\end{displaymath}
where: $n$ is the number of colliding bunches; $N_{1,2}$ is the number of protons per bunch in the two beams; $f_{\rm{rev}}$ is the accelerator revolution frequency; \hbox{$R(0,0)$} is the head-on rate; $S_{\rm{X}}$ and $S_{\rm{Y}}$ are the scan areas, defined as the area below the \hbox{$R(\Delta X,0)$} curve and the area below the \hbox{$R(0, \Delta Y)$} curve, respectively. In the assumption that the beam profiles are gaussian, the scan area can be obtained via a fit. However, the gaussian assumption is not necessary for the validity of the method: thus, the scan areas can simply be obtained via numerical integration.
The cross section $\sigma_{\rm{R}}$ for the chosen reference process can be obtained as \hbox{$\sigma_{\rm{R}} = R(0,0)/L$}. 

\section{$\textbf{pp}$ collisions}

\begin{figure}[!hbtp]
   \centering
           \includegraphics[width=0.5\textwidth, height = 0.39\textwidth]{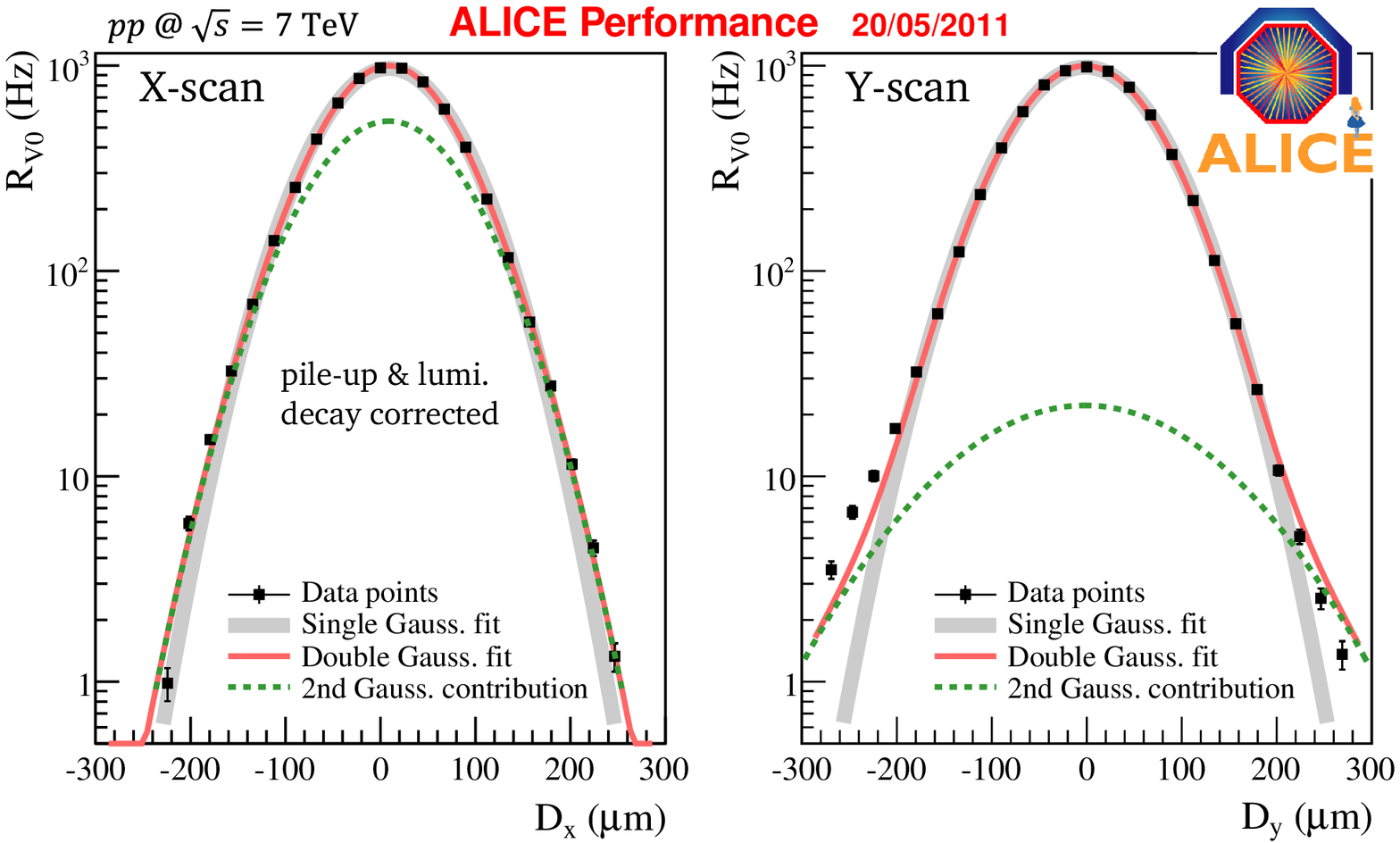}
	   \includegraphics[width=0.5\textwidth, height = 0.39\textwidth]{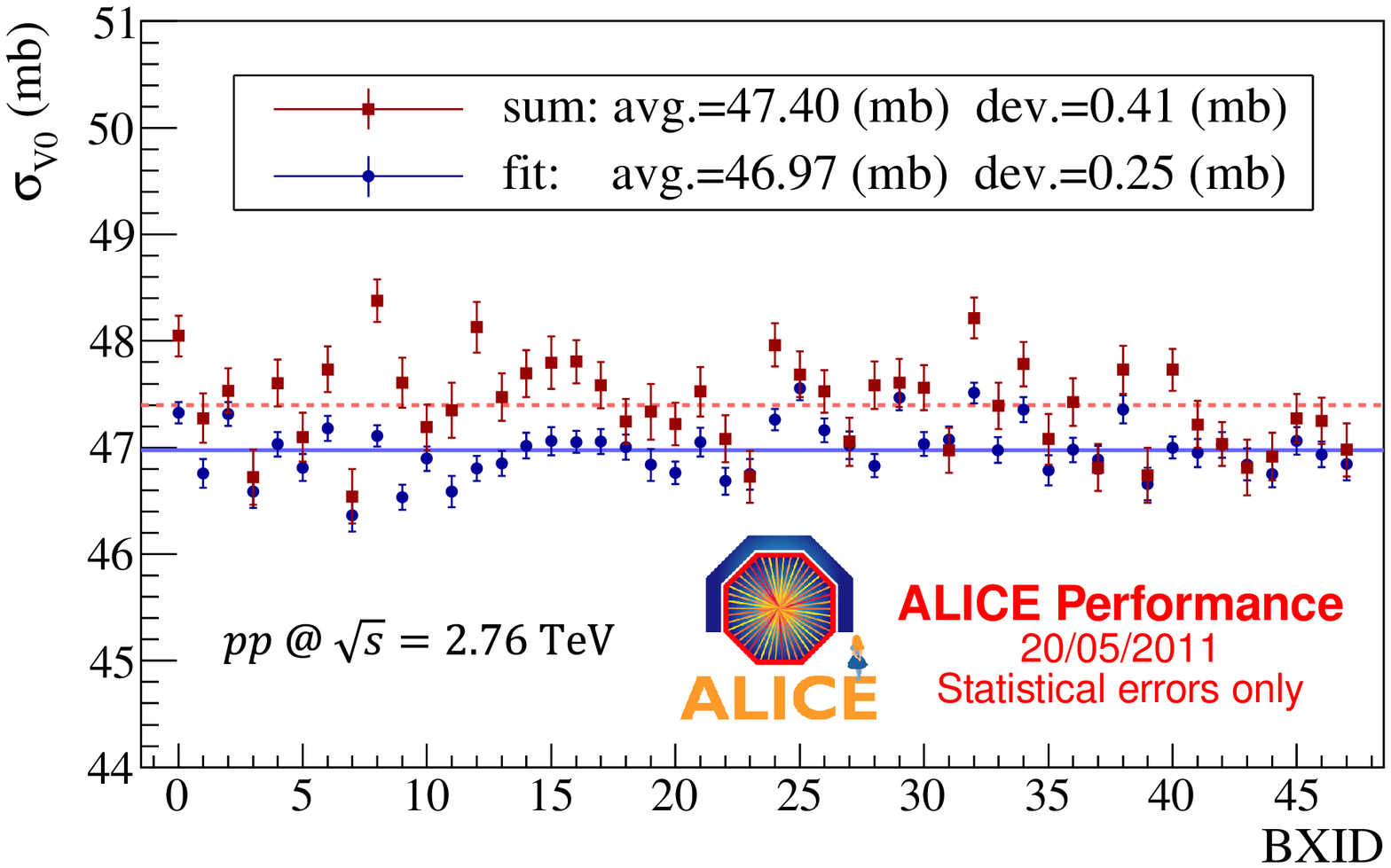}
     \label{fig:scan}      
   \caption{Left: VBand trigger rate vs beam separation in X and Y obtained during the May 2010 van der Meer scan. Gaussian and double Gaussian fits to the data are also shown. Right: Measured VBand cross section for 48 colliding bunch pairs in the March 2011 scan. Fit and numerical sum methods are compared.}
\end{figure}

In this section, we report about three pp scans carried out at the LHC\footnote{In this Conference, final results were presented for the May 2010 scan, together with preliminary results for the March 2011 scan. Results from the October 2010 scan were not presented. Since then, analysis for the October 2010 scan has been finalised: the results are reported in these proceedings.}. The colliding system details for all scans are specified in Table~\ref{tab:scans}. 
\begin{table}
\begin{tabular}{lccc}  
 Scan  & May 2010  & October 2010 &  March 2011 \\  \hline
 $\sqrt{s}$ (TeV) & 7 & 7 &  2.76 \\ 
 Colliding bunches in ALICE & 1x1 & 1x1  & 48x48 \\ 
  $\sigma_{\rm{VBand}}$ (mb) & 54.2$\pm$3.8 & 54.1$\pm$2.1 & \textit{47.4$\pm$3.3} \\ \hline
  Uncertainties  &   &  &   \\  
 Beam intensity & 4.4\% & 3.1\% &  \textit{5\%}  \\ 
 Length scale & 2.8\% & 1.4\% & \textit{2.8\%}  \\
 Luminosity decay   & 1\% & 1\% & \textit{1\%}  \\
 Beam centering & 1\% & 0.7\% & \textit{1\%} \\
 Satellite population & negligible & 1\% & negligible  \\
 Same fill reproducibility & n.a. & 1.2\% & n.a. \\
 Different fill reproducibility & 2.5\% & dropped & \textit{2.5\%}  \\
 Gaussian fit vs numerical sum & 2\% & dropped & \textit{2.5\%}  \\ 
 Total & 7\% & 4\% & \textit{7\%}  \\
\end{tabular}
\caption{Colliding system details, measured VBand cross sections and uncertainty sources for three pp vdM scans at the LHC. Values in italic are preliminary. The different fill reproducibility uncertainty assigned to the May 2010 result originates from the discrepancy with the preliminary October 2010 result \cite{Ken} (now obsolete).}
\label{tab:scans}
\end{table}
The chosen reference process (VBand) for all scans is the coincidence of hits in the V0 detector \cite{alice_JINST}, which consists of two scintillator arrays located on opposite sides of the interaction point (IP): \hbox{2.8<$\eta$<5.1} (V0A) and \hbox{-3.7<$\eta$<-1.7} (V0C). The VBand contamination from beam-gas collisions and afterpulses during the scan was measured by means of dedicated bunch crossing masks and found to be on the order of 0.1\%. 
The VBand rate was measured as a function of the beam separation (Fig.~\ref{fig:scan}). The scan areas have been obtained with the numerical sum method.
A set of corrections must be applied to the measured rate, namely: pile-up correction
(up to 40\%); length scale calibration, needed for a precise determination of the beam separation and performed by displacing the beams in the same direction and measuring the primary vertex displacement with a pixel detector (SPD \cite{alice_JINST}); satellite (displaced) collisions of protons captured in non-nominal RF slots, spotted via the arrival time difference in the two V0 arrays \cite{bcnwg2}; beam intensity decay and emittance growth in time; beam centering, accounting for the fact that beams may not be completely overlapping in one transverse direction while scanning the other one.
The beam intensity is measured separately for each circulating bunch by the LHC beam current transformers, and provided to the experiments after detailed analysis \cite{bcnwg2,bcnwg1}. 

In October 2010, two scans were performed in the same fill, in order to check the reproducibility of the measurement. The two results agree within 1.2\%: they have been averaged and the discrepancy added to the systematic uncertainties.   
In the March 2011 scan, 48 bunch pairs were colliding in ALICE. The cross section was measured separately for all bunch pairs (Fig.~\ref{fig:scan}), and the results averaged\footnote{The spread of cross sections measured with different bunches is less than 2\%, and the RMS is 0.8\%.}. 
 The results obtained in the three scans are presented in Table~\ref{tab:scans}. The various sources of systematic uncertainty are listed in Table~\ref{tab:scans}. 
With the availability of more scans, our level of confidence in the results has increased and some of the systematic uncertainties included in the May 2010 (and preliminary March 2011) result have been dropped from the October 2010 result. This fact, together with the improvement of the uncertainty on the beam intensities, led to a reduction of the total systematic uncertainty from 7\% to less than 4\% from May 2010 to October 2010. 

The ALICE luminosity determination has been compared to other LHC experiments via the cross section for a candle process: at least one charged particle produced with $p_{\rm{T}}$>0.5~GeV/c and |$\eta$|<0.8. The ALICE result\footnote{The result is based on the May 2010 vdM scan. The beam current uncertainty has been dropped since it is mainly correlated between LHC experiments.} for such cross section is   
\hbox{$\sigma_{\rm{candle}}$=0.78$\cdot \sigma_{\rm{VBand}}$=(42.4$\pm$2.0)~mb}, in good agreement with the ATLAS and CMS results \cite{heinemann}.

\section{$\textbf{Pb--Pb}$ collisions}

A vdM scan with Pb beams at the energy of 1.38~TeV per nucleon was performed in November 2010. 114 bunches were colliding in ALICE. A few reference processes were scanned, using the ALICE Zero Degree Calorimeters (ZDC \cite{alice_JINST}): such system consists of two neutron calorimeters (ZNA and ZNC) located on opposite sides of the IP, 114~m away from it, and of two proton calorimeters, not used in this analysis. The system is completed by two small electromagnetic calorimeters (ZEM), located only on one side, 7.5~m away from the IP. The analysis technique is the same as described for the pp case. Scanned processes and preliminary results for their cross sections are summarised in Table~\ref{tab:PbScan}. The main source of uncertainty is the fraction of ghost charge in the measured beam current, consisting of ions circulating along the LHC rings outside of nominally filled bunch slots, which do not contribute to the luminosity. Ghost charge corrrection is not applied to the result but a preliminary upper limit of 11\% (at 68\% confidence level) has been set \cite{bcnwg_priv}. Other uncertainties arise from the analysis procedure ($\simeq$5\%) and from the beam current transformers \cite{bcnwg_priv} ($\simeq$3\%), summing up to a total systematic uncertainty of -6\%+12\%.  
\begin{table}
\begin{tabular}{lcc}  
  Process &  Trigger & Cross section (b) \\  \hline
 ZNor  & ZNA || ZNC &     $363_{-22}^{+44}$ (syst.)        \\
 ZNand\_m & ZNA \&\& ZNC \&\& !ZEM  & $7.1\pm0.2$(stat.)$_{-0.4}^{+0.8}$ (syst.)\\
 ZNand\_h & ZNA \&\& ZNC \&\& ZEM & $5.9\pm0.2$(stat.)$_{-0.4}^{+0.7}$ (syst.) \\
\end{tabular}
\caption{Scanned processes and preliminary results for the November 2010 vdM scan with Pb beams.}
\label{tab:PbScan}
\end{table}

\section{Selected applications of vdM scans in ALICE}

\subsection{Measurement of the pp inelastic cross section}

\begin{figure}[!hbtp]
   \centering
           \includegraphics[width=0.35\textwidth]{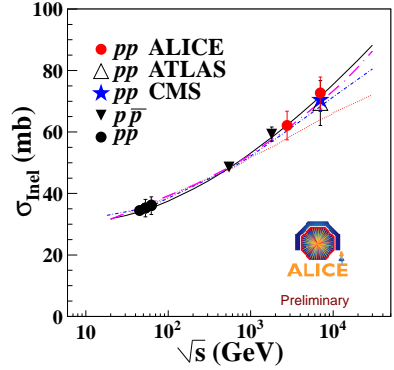}
	   \label{fig:inel}      
   \caption{Inelastic pp cross section as a function of $\sqrt{s}$: the ALICE measurements are compared with theoretical predictions \cite{inel1} and data from other  pp and $\rm{p\overline{p}}$ experiments \cite{inel2} (ISR, UA4, CDF, E811, ATLAS, CMS).}
\end{figure}

The VBand cross sections measured in the May 2010 and March 2011 van der Meer scans have been used to measure the pp inelastic cross section at \hbox{$\sqrt{s}$ = 7} and 2.76~TeV, respectively. The relative rates of single- and double- diffractive processes were measured \cite{martin} by studying properties of gaps in the pseudorapidity distribution of produced particles. Monte Carlo simulation tuned with the obtained ratios was used to determine the VBand efficiency for detecting inelastic pp interactions. Such efficiency was found to be  (75$\pm$1)\% at 7~TeV and (76$\pm$2)\% at 2.76~TeV. The resulting inelastic cross sections are \hbox{72.7$\pm$1.1(MC)$\pm$5.1(vdM) mb} at 7 TeV and \hbox{62.1$\pm$1.6(MC)$\pm$4.3(vdM) mb} at 2.76~TeV.
Such preliminary results are in agreement with preliminary results from other experiments and theoretical predictions (Fig.~\ref{fig:inel}).

\subsection{Measurement of the Pb--Pb electromagnetic dissociation cross sections}

\begin{figure}[!hbtp]
   \centering
           \includegraphics[height=0.32\textwidth,width=0.4\textwidth]{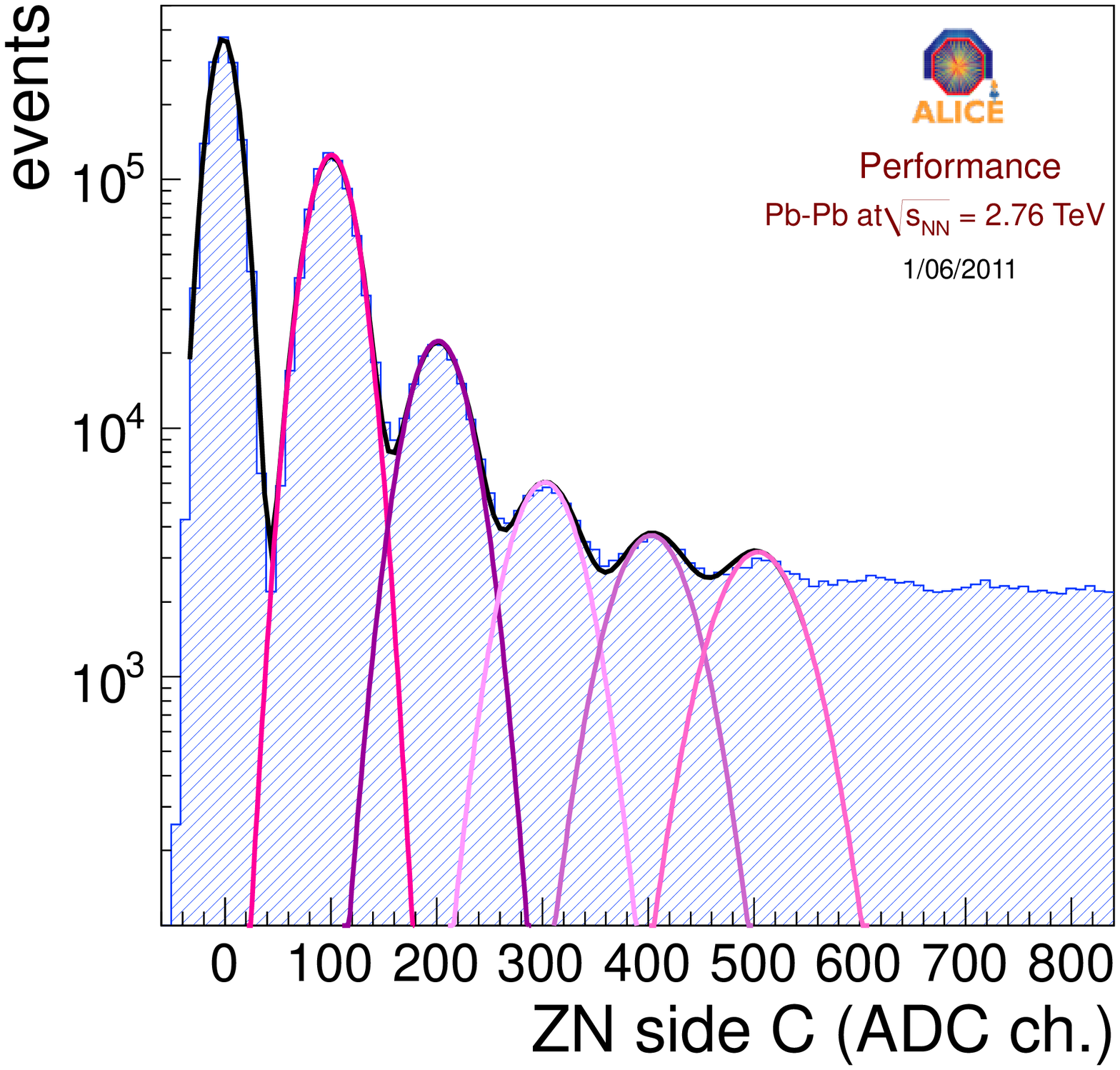}
	   \label{fig:zn}      
   \caption{Energy deposition spectrum in one of the ALICE neutron calorimeters, in a sample of ZNor-triggered events, fitted with a convolution of gaussian functions.}
\end{figure}

Electromagnetic dissociation \cite{Jackson} of heavy nuclei in ultra-peripheral collisions is one of the main sources of beam loss in heavy-ion accelerators. The neutrons emitted in the process have a rapidity very close to the beam one, and can then be detected with the Zero Degree Calorimeters. The single and mutual electromagnetic dissociation cross section have been measured \cite{chiara} from a sample of events triggered by the ZNor \hbox{(ZNA || ZNC)} process. The fractions of events in the sample where a given number of neutrons is emitted can be obtained via a fit to the energy deposition spectra in the calorimeters (Fig.~\ref{fig:zn}). The fraction of events with at least one neutron in one of the two ZNs is proportional to the sum of the cross sections for single electromagnetic dissociation ($\sigma_{\rm{sEMD}}$) and for hadronic interaction ($\sigma_{\rm{had}}$), while the fraction of events with at least one neutron in one ZN and nothing in the other one is proportional to the difference between $\sigma_{\rm{sEMD}}$ and the mutual electromagnetic dissociation cross section ($\sigma_{\rm{mEMD}}$). In both cases, the  constant of proportionality is the trigger cross section $\sigma_{\rm{ZNor}}$, measured in the November vdM scan and reported in Table~\ref{tab:PbScan}.
\begin{table}
\begin{tabular}{lcc}  
   & ALICE preliminary  & RELDIS \\  \hline
 $\sigma_{\rm{sEMD}}$ (b)  & $186_{-11}^{+23}$(syst.) &     $185\pm9$ (syst.)        \\
 $\sigma_{\rm{mEMD}}$ (b)  & $5.7\pm0.2$(stat.)$_{-0.3}^{+0.7}$(syst.) & $5.5\pm0.6$(syst.)\\
\end{tabular}
\caption{ALICE preliminary results for the Pb nuclei electromagnetic dissociation cross section, compared to the RELDIS predictions.}
\label{tab:emd}
\end{table}
The measured cross sections $\sigma_{\rm{ZNand\_m}}$ and $\sigma_{\rm{ZNand\_h}}$ are linear combinations of the hadronic and mutual  electromagnetic dissociation cross sections: the coefficients, related to the ZEM efficiency for detecting mutual electromagnetic and hadronic interactions, have been estimated by simulation using HIJING and RELDIS \cite{reldis} for the hadronic and the electromagnetic part respectively. Using all the above information, $\sigma_{\rm{sEMD}}$ and $\sigma_{\rm{mEMD}}$ can be extracted: they are reported in Table~\ref{tab:emd}, compared to the RELDIS predictions, which are found to be in good agreement with our results.

\section{Conclusions}

Reference cross sections for luminosity normalisation have been measured by the \hbox{ALICE} experiment in van der Meer scans at the LHC, in both pp and Pb--Pb collisions. The achieved precision is as good as 4\% in pp collisions. Such measurements allowed ALICE to obtain preliminary results for physical observables such as the pp inelastic cross section and the Pb--Pb electromagnetic dissociation cross sections.







\bibliographystyle{aipproc}   




\end{document}